\documentstyle[aas2pp4,psfig,11pt]{article}

\begin{document}

\title{Discovery of twin kHz quasi-periodic oscillations in the high galactic latitude X-ray transient XTE J2123--058}

\author{Jeroen Homan\altaffilmark{1},
        Mariano M\'endez\altaffilmark{1,2},
        Rudy Wijnands\altaffilmark{1},
        Michiel van der Klis\altaffilmark{1},
        Jan van Paradijs\altaffilmark{1,3}}

\altaffiltext{1}{Astronomical Institute 'Anton Pannekoek', University of Amsterdam, and Center for High Energy Astrophysics,  Kruislaan 403, 1098 SJ, Amsterdam, The Netherlands}
\altaffiltext{2}{Facultad de Ciencias Astron\'omicas y Geof\'{\i}sicas, Universidad Nacional de La Plata, Paseo del Bosque S/N, 1900 La Plata, Argentina}
\altaffiltext{3}{University of Alabama in Huntsville, Department of Physics, Huntsville, AL 45899}

\begin{abstract}
We report the discovery of twin kHz quasi-periodic oscillations (QPOs)
in the persistent X-ray flux of the new X-ray transient \mbox{XTE
J2123--058}. We detected the QPOs in data taken with the proportional
counter array on board the {\it Rossi X-ray Timing Explorer} on 1998
July 14, when the source count rate was at about 50\% of its outburst
peak level. The frequencies of the QPOs were 853$\pm$4 and 1129$\pm$8
Hz. The peaks had widths of 32$^{+11}_{-8}$ and 50$^{+17}_{-13}$ Hz
and the rms amplitudes were 6.1$^{+0.7}_{-0.6}$\% and
6.5$^{+0.9}_{-0.8}$\%, respectively. The QPO frequencies increased
marginally with count rate and the amplitudes showed a small increase
with photon energy. On the basis of the appearance in the color-color
and hardness-intensity diagrams, the low-frequency power spectrum, and
the strength of the kHz QPOs, the system can be classified as an atoll
source. The source showed five type I X-ray bursts. The distance to
the source is estimated to be about 10 (5--15) kpc, which combined
with the high Galactic latitude of 36$^{\circ}$.2 indicates the source
is a located in the Galactic halo.
\end{abstract}


\section{Introduction}
The X-ray transient XTE J2123--058 was discovered with the all-sky
monitor (ASM; Levine et al. \markcite{lebrcu96}1996) on board the {\it
Rossi X-ray Timing Explorer} ({\it RXTE}; Bradt, Rothschild, \& Swank
\markcite{brrosw93}1993) on 1998 June 27 (Levine, Swank, \& Smith
\markcite{leswsm98}1998). Subsequent observations with the
proportional counter array (PCA, also on board {\it RXTE}; Jahoda et
al. \markcite{jaswgi96}1996) showed two weak X-ray bursts (Takeshima
\& Strohmayer \markcite{tast98}1998), probably of type I (i.e.,
thermonuclear origin), indicating that the compact object is a neutron
star. Observations of the optical counterpart (Tomsick et
al. \markcite{tohale98}1998a) revealed an orbital period of
5.9567$\pm$0.0033 hr (Tomsick et al. \markcite{tokeha98}1998b;
Ilovaisky \& Chevalier \markcite{ilch98}1998). The source has a high
galactic latitude of 36$^{\circ}$.2, which is unusual for an X-ray
transient.

Quasi-periodic oscillations (QPOs) with frequencies between 200 and
1200 Hz, the so called kHz QPOs (Van der Klis \markcite{kl98}1998),
have been found in the persistent emission of numerous low-mass X-ray
binaries (LMXBs). They are often found in pairs, with a frequency
difference ranging from 250 to 350 Hz. The strength of the
oscillations depends on the type (Hasinger \& Van der Klis
\markcite{hakl89}1989) of LMXB; they are usually stronger in atoll
sources than in Z sources. A few sources have shown oscillations
during type I X-ray bursts (e.g., Strohmayer et
al. \markcite{stzhsw96}1996, \markcite{stzhsw98}1998), whose
frequencies are believed to be close to once or twice (Miller
\markcite{mi98}1998) the neutron star spin frequency, and which are
near the kHz QPO frequency difference. On this basis it has been
suggested (Strohmayer et al. \markcite{stzhsw96}1996; see also Miller,
Lamb, \& Psaltis \markcite{milaps98}1998) that the frequency
difference is equal to the neutron star spin frequency, but this idea
has been seriously challenged by results by Van der Klis et
al. \markcite{klwiho97}(1997), M\'endez et
al. \markcite{meklfo98}(1998a), and M\'endez, Van der Klis, \& Van
Paradijs \markcite{meklpa98}(1998).

In this Letter, we present the discovery of kHz QPOs in XTE
J2123--058. Preliminary results of this work were already presented by
Homan et al. \markcite{hoklpa98}(1998).

\section{Observation and Analysis}
For our analysis, we used data obtained with the PCA. Data were taken
on five occasions (see Table 1) for a total of $\sim$43 ks. PCA
detectors 4 and 5 were inactive during observation 1, detector 5 was
inactive during the last $\sim$1000 s of observation 3, and detector 4
for $\sim$13000 s during observation 4. Observation 2 consisted of two
scans over the region to determine a better source position (Takeshima
\& Strohmayer \cite{tast98}1998); data from that observation were not
included in the color diagrams and only $\sim$100 s were used for
timing analysis.  Apart from the Standard 1 (0.125 s time resolution
in one energy channel) and Standard 2 modes (16 s time resolution in
129 energy bands), which were always active, data were obtained in an
Event modye with a $2^{-13}$ s time resolution and 64 energy bands
(observations 2 and 3), and in a Good Xenon mode with a $2^{-20}$ s
time resolution and 256 energy bands (observations 4 and 5). All modes
covered the 2--60 keV energy band. Observations 1 did not yield data
with the time resolution required to search for kHz QPOs.

The Standard 2 data of detectors 1--3 were background corrected, and
used to create light curves, color-color diagrams (CDs), and
hardness-intensity diagrams (HIDs). Figure 1 shows the ASM and PCA
lightcurves. For observation 2 only the maximum count rate reached
during the two dwells is plotted (open circle). Figure 2 shows the CD
and HID of observations 1 and 3--5 combined. Data during X-ray bursts
were excluded from Figures 1 and 2.

Using all available high time resolution data we made Fourier
transforms of 16 s data segments to create power spectra ranging from
1/16 to 2048 Hz to search for kHz QPOs. To study the low-frequency
power spectrum we also created 1/128--512 Hz power spectra using 128 s
data segments. Power spectra were selected on time, count rate, and
position in the CD and HID, and averaged before further analysis. The
properties of the kHz QPOs were measured by fitting the resulting
average power spectra from 300--2048 Hz with a fit function comprised
of a constant, representing the Poisson level, and a Lorentzian for
each QPO. The low-frequency power spectra were fitted with a power law
plus a power law with an exponential cutoff. The values for the
rms amplitudes are background corrected. Errors on the fit parameters
were determined using $\Delta\chi^2=1$.  All the upper limits quoted
are 95\% confidence limits. In case of the kHz QPOs upper limits were
determined by fixing the full-width-at-half-maximum (FWHM) to 25 Hz.

\begin{table}[t]
\begin{tabular}{cccc}
Obs. & Date & Begin (UTC) & End (UTC) \\
\hline 
\hline
1 & 27-06-1998 & 23:29 & 23:42 \\
2 & 29-06-1998 & 00:33 & 01:09 \\
3 & 29-06-1998 & 16:01 & 17:20 \\
4 & 14-07-1998 & 09:41 & 19:42 \\
5 & 22-07-1998 & 04:54 & 17:49 \\
\hline
\end{tabular}
\caption{{\footnotesize Log of the PCA observations of XTE J2123--058.}}
\end{table}

\begin{figure}[t]
\psfig{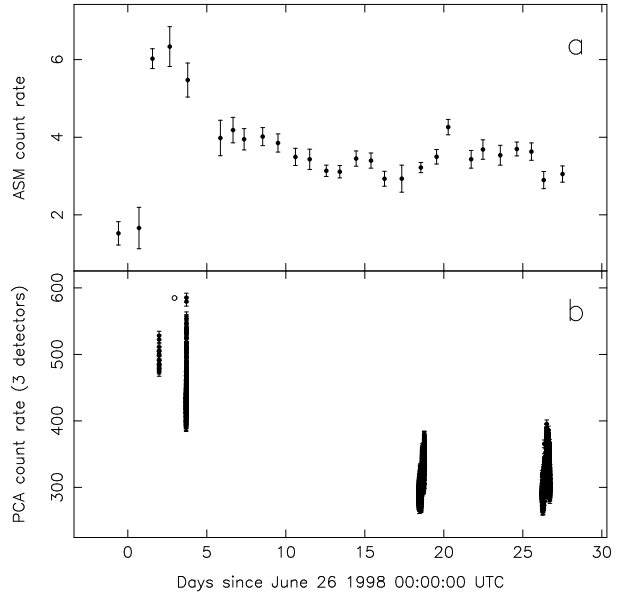}
\caption{{\footnotesize The ASM (a) and PCA (b) lightcurves of XTE J2123--058. The ASM (1.5--12 keV) points are one day averages, the PCA points 16 s averages. For the PCA (2--60 keV) light curve we only used background corrected count rates from the first three detectors. The open circle depicts the maximum count rate reached during the dwells of observation 2.}}
\end{figure}

\begin{figure}[t]
\psfig{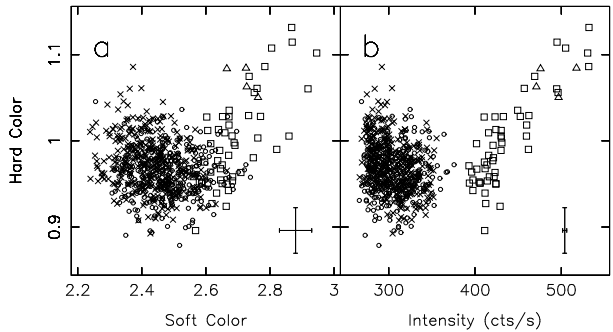}
\caption{{\footnotesize The color-color (a) and hardness-intensity (b) diagrams of observations 1 (triangles), 3 (squares), 4 (crosses), and 5 (circles). The soft color is the ratio of the count rates in the 3.5--6.4 and 2--3.5 keV bands and the hard color the ratio of the count rates in the 8.6--16.0 and 6.4--8.6 keV bands. The intensity is the count rate in the 2--16.0 keV band. The data points are 64 s averages. Typical error bars are shown.}}
\end{figure}

\begin{figure}[t]
\psfig{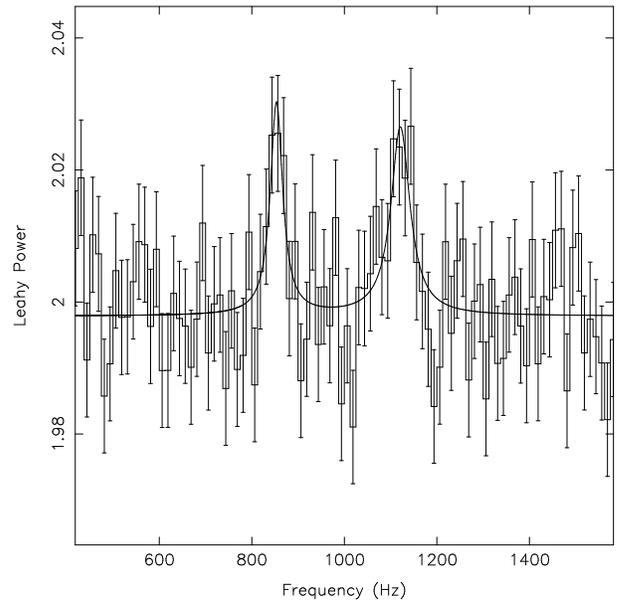}
\caption{{\footnotesize The power spectrum of the first two orbits of observation 4 combined, showing the two kHz QPOs.}}
\end{figure}

\begin{figure}[t]
\psfig{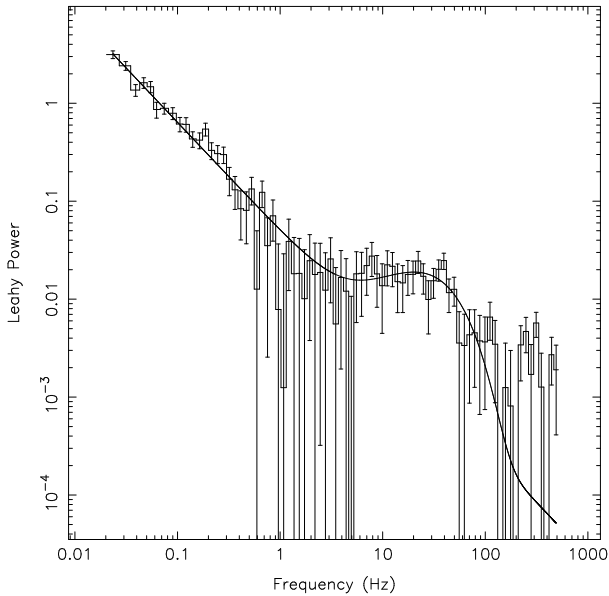}
\caption{{\footnotesize The average broad-band power spectrum of observations 3, 4, and 5 combined. The Poisson level has been subtracted.}}
\end{figure}

\section{Results}

We found two simultaneous kHz QPOs in the data of observation 4 (see
Figure 3). This observation consisted of six $\sim$3300 s intervals of
contiguous data, hereafter orbits, separated by $\sim$2400 s intervals
during which no data were obtained. The QPOs were strongest during the
first two orbits, when the count rate was lowest. In these orbits the
QPOs had frequencies of 853$\pm$4 Hz and 1129$\pm$8 Hz, widths of
32$^{+11}_{-8}$ and Hz 50$^{+18}_{-13}$ Hz, and rms amplitudes of
6.1\%$^{+0.7}_{-0.6}$\% and 6.5\%$^{+0.9}_{-0.8}$\% (unless otherwise
stated all fit parameters are those measured in the 2--11.2 keV
band). The mean 2--60 keV background-corrected count rate during the
first two orbits was 485 counts s$^{-1}$.

There are indications for an increase of QPO strength with photon
energy. In the 2--5.7 keV band only upper limits could be determined
of 6.2\% and 5.2\% rms, for the lower and higher frequency QPO,
respectively. In the 5.7--11.2 keV band the frequencies and widths are
consistent with the values obtained in the 2--11.2 keV band; the
amplitudes are 8.8\%$^{+1.2}_{-1.1}$\% and 9.0\%$^{+1.5}_{-1.3}$\%
rms. Only upper limits could be determined in the 11.2--16.7 keV band
of 18.5\% and 24.2\% rms.

A more detailed analysis showed a difference between the data of the
first and second orbit. The power spectra of the first orbit clearly
showed the low frequency QPO at 849$\pm$2 Hz with a width of
20$^{+7}_{-5}$ Hz and an amplitude of 6.7\%$\pm$0.7\% rms, whereas the
other QPO was only marginally detectable at 1110$\pm$10 Hz with an
amplitude of 4.8\%$\pm$0.9\% rms and its width fixed at 35 Hz. In the
power spectra of the second orbit the lower QPO is very narrow and its
width was fixed at 6 Hz; it had a frequency of 871$\pm$2 Hz and an
amplitude of 3.4\%$^{+1.2}_{-0.7}$\% rms. The high frequency QPO was
found at 1141$^{+4}_{-5}$ Hz with a width of 32$^{+15}_{-11}$ Hz and
an amplitude of 7.1\%$^{+1.0}_{-0.6}$\% rms. The mean (2--60 keV)
background corrected count rates of the first and second orbits were,
respectively, 478 and 497 counts s$^{-1}$, with standard deviations
(for 16s time bins) of 22 and 16 counts s$^{-1}$. So, the QPO
frequency increased with count rate while the frequency difference was
consistent with being constant at $\sim$265 Hz.

For the last four orbits of observation 4, and for the other
observations, only upper limits on the presence of kHz QPOs could be
determined. They were, in the 2--11.2 keV band, 16\%, 4.4\%, 4.3\%,
and 4.5\% rms for, respectively, observation 2, 3, 4 (last four
orbits), and 5.

Selections of all the observation 4 data based on the position in the
CD and HID showed that the QPO are only found at the lowest soft
colors and count rates. No other significant dependence of QPO
properties on color or intensity could be distinguished based on these
selections. 

We performed time lag measurements between the 3.5--9.3 keV and
9.3--14.9 keV energy bands. In both QPOs, the lags were consistent
with being zero, but also with what has been found by Vaughan et
al. \markcite{vaklme97}(1997, 1998) for other sources with kHz
QPOs. The lags are $-8\pm4\times10^{-2}$ ms for the lower and
$-5\pm4\times10^{-2}$ ms for the higher frequency QPO, where the minus
sign indicates a soft lag.

Observations 3, 4, and 5 were combined to study the low frequency
power spectrum (see Fig. 4). The average power spectrum (2--11.2
keV) was fitted with a power law plus a power law with an exponential
cutoff. The rms amplitude of the power law component was
2.61\%$\pm$0.05\% (integrated over 0.01--1 Hz) and its power law index
was 1.10$\pm$0.04. The second component had an rms amplitude of
4.5\%$\pm$0.3\% (1--100 Hz), a power law index of --1.3$\pm$0.5, and a
cut-off frequency of 17$\pm$5 Hz. Selections were made on soft
color. For soft colors higher than 2.6 the average power spectrum is
well described by a single power law with an rms amplitude of
2.40\%$\pm$0.15\% and a power law index of 1.23$\pm$0.15. For a possible
power law component with an exponential cutoff an upper limit of 2\%
was determined. The average power spectrum for soft colors lower than
2.6 was fitted with a power law (2.47\%$\pm$0.05\% rms and power law
index of 1.07$\pm$0.04) plus an exponentially cut-off power law
(5.1\%$\pm$0.3\% rms, power law index of --1.2$\pm$0.4, and a cut-off
frequency of 20$\pm$6 Hz). In the power spectra of the last four
orbits of observation 4, an equally good fit could be obtained by
replacing the cut-off power law by a Lorentzian at 39$\pm$2 Hz, with
an rms amplitude of 5.0\%$\pm$0.6\% and a FWHM of 20$\pm$7 Hz.

We have analyzed five X-ray bursts which, based on the spectral
softening during their decay, are most probably of type I . We
searched for burst oscillations in 1s power spectra (1--1024 Hz). This
was done in the 2--60 keV and 3.5--15 keV energy bands, during rise
and decay of each burst. Data of the five bursts were also added to
increase sensitivity. No oscillations were found, with upper limits on
their amplitudes of 20\% rms. Two very strong bursts were observed
during observation 5. When corrected for the persistent emission
obtained from the 100 s just before the bursts, their peak count rates
(2--11.2 keV) are 2550$\pm$60 and 2540$\pm$60 counts s$^{-1}$. These
peak count rates are very close to each other and also much higher
than those of the other three bursts (which had peak count rates of
$\sim$500, $\sim$1200 and $\sim$330 counts s$^{-1}$). It is possible
that the peak count rates are so similar because they are related to
the Eddington luminosity reached in photospheric radius expansion
bursts. Indeed both bursts show a splitting of their main peak at
higher photon energies, which usually is an indication for radius
expansion (Lewin, Van Paradijs, \& Taam,
\markcite{lepata93}1993). Fitting the burst spectra with a blackbody
model did not conclusively show that radius expansion had occurred.
However, an anti-correlation was found between the temperature and the
radius of the blackbody, a feature that is typical for radius
expansion bursts (Lewin, Van Paradijs, \& Taam,
\markcite{lepata93}1993). If the bursts are indeed radius expansion
bursts, the distance to the source can be estimated (see Lewin, Van
Paradijs, \& Taam, \markcite{lepata93}1993). Assuming cosmic
abundancies, a 1.4 $M_\odot$ neutron star mass, 10 km radius, and
isotropic emission, the upper limit to the distance is $\sim$14
kpc. Using the observational relation between $M_V$, $P_{orb}$, and
$L_X/L_{Edd}$ found by Van Paradijs \& McClintock
\markcite{pamc}(1994) we can estimate the distance in another
way. Taking $m_V$=16.8 (Tomsick et al. 1998b), $A_V=0.36$ (Schlegel,
Finkbeiner, \& Davis, \markcite{scfida98}1998),
$F_X\approx2.5\times10^{-9}$ erg s$^{-1}$ (from observation 3, which
was closest to the optical observation), and the orbital period found
by Tomsick et al. \markcite{tokeha98}(1998b) and Ilovaisky \&
Chevalier \markcite{ilch98}(1998) we find a distance to the source
between 10 and 15 kpc. Finally, a distance estimate can be made on the
basis of the average burst duration ($\tau$) and/or the observed ratio
of persistent X-ray flux to the average flux emitted in X-ray bursts
($\alpha$), both of which show a strong correlation with
$\gamma=L_X/L_{Edd}$ (Lewin, Van Paradijs, \& Taam,
\markcite{lepata93}1993). For observation 5 ($\tau\approx$12 s,
$\alpha\approx$120, and $F_X\approx1\times10^{-9}$ erg s$^{-1}$) this
gives a distance range of 5 to 11 kpc. Combining the three methods
gives a distance that is probably about 10 (5--15) kpc, which, using
the galactic latitude of --36$^{\circ}$.2, gives a distance from the
Galactic plane of 6 (3--9) kpc.
 
\section{Discussion}
We have found two simultaneous kHz QPOs in the X-ray flux of XTE
J2123--058. The frequencies of the QPOs, $\sim$850--870 Hz for the
lower and $\sim$1110--1140 Hz for the upper peak, are within the range
observed for other sources. Their widths and rms amplitudes are also
comparable to what is found in other LMXBs (Van der Klis,
\markcite{kl98}1998). The QPOs are only found when the count rate and
soft color are lowest. When the QPO frequencies changed $\sim$20--30
Hz, the frequency difference remained consistent with being constant
at $\sim$265 Hz.

From the other sources showing kHz QPOs it was already evident that
the QPO frequencies, although strongly dependent in a given source on
the difference between the instantaneous luminosity and the average
luminosity, do not depend on the average luminosity itself.  XTE
J2123--058 complicates this puzzle. A transient with a time averaged
luminosity a factor $\sim$10--2000 (averaging the total outburst
luminosity over 3 years) lower than in the persistent sources, it
again shows exactly the same kHz QPO frequencies.

Several properties (Hasinger \& Van der Klis \markcite{hakl98}1989)
suggest that XTE J2123--058 is an atoll source. (1) The appearance in
the CD and HID is reminiscent of the shapes traced out by atoll
sources; the pattern traced out by XTE J2123--058 is then probably
related to the lower and upper banana. (2) The low-frequency power
spectra can be fitted with a power law component (which we identify
with very low frequency noise) plus an exponentially cut-off power law
 (which we identify with high frequency noise). As in atoll
sources the high-frequency noise is much weaker when the source is in
the upper banana than in the lower banana. Also, the strength of these
noise components is comparable to that found in atoll sources on the
banana branch (Van der Klis, \markcite{kl95}1995). (3) The strength of
the kHz QPOs is relatively high ($\sim$9\%); a common value for atoll
sources. On this basis we classify this source as an atoll source. The
burst properties strengthen this classification. At least five type-I
X-ray burst were observed during the observations. Although there are
two Z sources (Cyg X-2 and GX 17+2) that occasionally show X-ray
bursts, the high burst frequency (five in $\sim$43 ks) is
usually regarded as a property of atoll sources.  From the two strong
bursts we can derive a value for the persistent luminosity that is
smaller than $\sim$0.35$L_{Edd}$, which is typical for atoll sources.
The kHz QPOs are only seen in the lower banana, i.e., at low accretion
rates, which is also seen in other atoll sources (Wijnands et
al. \markcite{wiklpa97}1997; \markcite{wiklme98}1998; M\'endez et
al. \markcite{me98}1998b; Zhang et al. \markcite{zhsmst98}1998).

The estimated distance from the Galactic plane of 6 (3--9) kpc is
unusually high when compared to other LMXBs, and suggests that the
source is located in the Galactic halo. For 18 neutron star LMXBs Van
Paradijs \& White (\markcite{pawh95}1995) found an rms value for the
distance from the Galactic plane of 1.0 kpc. An important factor
determining the height distribution of LMXBs is the magnitude of any
kick received during the formation of the neutron star (Van Paradijs
\& White \markcite{pawh95}1995; Ramachandran \& Bhattacharya
\markcite{rabh97}1997) . These kicks can result in high system
velocities. We calculated the system velocity needed to reach a height
of 6 kpc, by integrating orbits in the gravitational potential of
Kuijken \& Gilmore \markcite{kugi}(1989). We found system velocities
between 150 and 400 km s$^{-1}$, depending on the pre-kick location in
the Galactic plane. These are relatively high velocities, and the
kicks needed to obtain them can easily disrupt a binary. Since the
distance of XTE J2123--058 from the Galactic plane is comparable to
that of globular clusters, another scenario might be that the source
was ejected from a globular cluster. In this case a much smaller
system velocity (typically 40 km s$^{-1}$) is required, which can
easily be obtained from binary--single star or binary--binary
interactions in a globular cluster (Phinney \& Sigurdsson
\markcite{phsi91}1991; Sigurdsson \& Phinney \markcite{siph95}1995).

\acknowledgments{This work was supported in part by the Netherlands
Foundation for Research in Astronomy (ASTRON) grant 781-76-017. MM is
a fellow of the Consejo Nacional de Investigaciones Cient\'{\i}ficas y
T\'ecnicas de la Rep\'ublica Argentina. JVP acknowledges NASA support
through grants NAG-5-4482 and NAG-5-7382. This research has made use
of data obtained through the High Energy Astrophysics Science Archive
Research Center Online Service, provided by the NASA/Goddard Space
Flight Center. We would like to thank the referee Keith Jahoda for his
helpful comments on the paper. We also thank Eric Ford for his
assistance with the data analysis, R. Ramachandran, Frank Verbunt and
Simon Portegies Zwart for their help and comments on the discussion
about the distance, and Peter Jonker for comments on an earlier
version of this document.}


{\footnotesize
\begin{references}
\reference{brrosw93} Bradt, H.V., Rothschild, R.E., Swank, J.H., 1993, A\&AS, 97, 355
\reference{hakl89} Hasinger, G. \& Van der Klis, M., 1989, \aap, 225, 79 
\reference{hoklpa98} Homan, J, Van der Klis, M., Van Paradijs, J., M\'endez, M., 1998, \iaucirc, 6971
\reference{ilch98} Ilovaisky, S.A., Chevalier, C., 1998, \iaucirc, 6957
\reference{jaswgi96} Jahoda, K., Swank, J.H., Giles, A.B., Stark, M.J., Strohmayer, T., Zhang, W., Morgan, E.H., 1996, SPIE, 2808, 59
\reference{kugi89} Kuijken, K., Gilmore, G., 1989, \mnras, 239, 571
\reference{lebrcu96} Levine, A.M., Bradt, H., Cui, W., Jernigan, J.G., Morgan, E.H., Remillard, R., Shirey, R.E., Smith, D.A., 1996, \aap, 469, L33
\reference{leswsm98} Levine, A., Swank, J., Smith, E., 1998, \iaucirc, 6955
\reference{lepata93} Lewin, W.H.G., Van Paradijs, J., Taam, R.E., 1993, Space Sci. Rev., 62, 223
\reference{meklfo98} M\'endez, M., Van der Klis, M., Ford, E.C., Van Paradijs, J., Vaughan, B.A., 1998a, \apjl, 505, 23
\reference{meklpa98} M\'endez, M., Van der Klis, M., Van Paradijs, J., 1998, \apjl, 506, 177
\reference{me98} M\'endez et al., 1998b, ApJ Letters, in press
\reference{mi98} Miller, C.M., 1998, submitted to ApJ (astro-ph 9809235)\reference{milaps98} Miller, M.C., Lamb, F.K., Psaltis, D., 1998, to appear in ApJ
\reference{phsi91} Phinney, E.S., \& Sigurdsson, S., 1991, \nat, 349, 220
\reference{rabh97} Ramachandran, R., Bhattacharya, D., 1997, \mnras, 288, 565
\reference{scfida98} Schlegel, D.J., Finkbeiner, D.P. \& Davis, M., 1998, ApJ, 500, 525S 
\reference{siph95} Sigurdsson, S., Phinney, E.S., 1995, \apjs, 99, 609
\reference{stzhsw96} Strohmayer, T.E., Zhang, W., Swank, J.H., Smale, A., Titarchuk, L., Day, C., 1996, \apjl, 469, L9
\reference{stzhsw98} Strohmayer, T.E., Zhang, W., Swank, J.H., White, N.E., Lapidus, I., 1998, \apjl, 498, 135 
\reference{tast98} Takeshima, T., Strohmayer, T.E., 1998, \iaucirc, 6958
\reference{tohale98} Tomsick, J.A., Halpern, J.P., Leighly, K.M., Perlman, E., 1998a, \iaucirc, 6957
\reference{tokeha98} Tomsick, J.A., Kemp, J., Halpern, J.P., Hurley-Keller, D., 1998b, \iaucirc, 6972
\reference{kl95} Van der Klis, M., 1995, in {\it X-Ray Binaries}, eds. Lewin, W.H.G., van Paradijs, J., van den Heuvel, E.P.J. Cambridge University Press, 252
\reference{klwiho97} Van der Klis, M., Wijnands, R.A.D., Horne, K., Chen, W., 1997, \apjl, 481, L97
\reference{kl98} Van der Klis, M. 1998, in 'The many faces of Neutron Stars', 
R. Buccheri, J. van Paradijs \& M.A. Alpar (Eds), 
NATO ASI Series C, Vol. 515, pp. 337-368 (Kluwer Academic Publishers)
\reference{pamc94} Van Paradijs, J., McClintock, J.E., 1994, A\&A, 290, 133
\reference{pawh95} Van Paradijs, J., White, N., 1995, \apj, 447, L33 
\reference{vaklme97} Vaughan, B.A., Van der Klis, M., M\'endez, M., Van Paradijs, J., Wijnands, R.A.D., Lewin, W.H.G., Lamb, F.K., Psaltis, D., Kuulkers, E.,  \& Oosterbroek, T., 1997, \apjl, 483, L115
\reference{vaklme98} Vaughan, B.A., Van der Klis, M., M\'endez, M., Van Paradijs, J., Wijnands, R.A.D., Lewin, W.H.G., Lamb, F.K., Psaltis, D., Kuulkers, E.,  \& Oosterbroek, T., 1998, \apjl, in press
\reference{wiklpa97} Wijnands, R., Van Der Klis, M., Van Paradijs, J., Lewin, W.H.G., Lamb, F.K., Vaughan, B.A., \& Kuulkers, E., 1997, \apj, 479, L141 
\reference{wiklme98} Wijnands, R., Van der Klis, M., M\'endez, M., Van paradijs, J., Lewin, W.H.G., Lamb, F.K., Vaughan, B., \& Kuulkers, E., 1998, \apj, 495, L39
\reference{zhsmst98}Zhang, W., Smale, A.P., Strohmayer, T.E, \& Swank, J.H., 1998, \apj, 500, L171
\end{references}}
\end{document}